\documentclass[runningheads]{llncs}
\usepackage[table,xcdraw]{xcolor}
\usepackage{graphicx}
\usepackage{cite}
\usepackage{adjustbox}
\usepackage{amsmath}
\usepackage[english]{babel}
\usepackage{float}
\usepackage{hyperref}
\usepackage[all]{hypcap}
\usepackage{svg}
\usepackage{tabularx}
\newcolumntype{Y}{>{\centering\arraybackslash}X}
\usepackage{booktabs}

\hypersetup{linktoc=all}

\AtBeginDocument{}

\begin{document}
\title{Efficient Pix2Vox++\ for 3D Cardiac Reconstruction from 2D echo views}
\titlerunning{3D Anatomy Reconstruction from 2D}
\author{
David Stojanovski \inst{1}
\thanks{
This work was supported by the Wellcome/EPSRC Centre for Medical Engineering [WT203148/Z/16/Z], by the British Heart Foundation [TG/17/3/33406] and the National Institute for Health Research (NIHR) Biomedical Research Centre at Guy's and St Thomas' NHS Foundation Trust and King's College London. Pablo Lamata holds a Wellcome Trust Senior Research Fellowship [209450/Z/17/Z]. The views expressed are those of the author(s) and not necessarily those of the NHS, the NIHR or the Department of Health. 
} \and
Uxio Hermida \inst{1}
\and
Marica Muffoletto \inst{1}
\and
Pablo Lamata \inst{1}
\and
Arian Beqiri \inst{1,2}
\and
Alberto Gomez \inst{1,2}
}

\authorrunning{Stojanovski et al.}

\institute{
 King’s College London, School of Biomedical Engineering \& Imaging Sciences,
London, SE1 7EU, UK \email{\{first.last\}@kcl.ac.uk}\\ \and
 Ultromics Ltd., Oxford, OX4 2SU, UK \email{\{first.last\}@ultromics.com}\\
 david.stojanovski@kcl.ac.uk}

\maketitle              
\begin{abstract}
Accurate geometric quantification of the human heart is a key step in the diagnosis of numerous cardiac diseases, and in the management of cardiac patients. Ultrasound imaging is the primary modality for cardiac imaging, however acquisition requires high operator skill, and its interpretation and analysis is difficult due to artifacts. Reconstructing cardiac anatomy in 3D can enable discovery of new biomarkers and make imaging less dependent on operator expertise, however most ultrasound systems only have 2D imaging capabilities. We propose both a simple alteration to the Pix2Vox++  networks for a sizeable reduction in memory usage and computational complexity, and a pipeline to perform reconstruction of 3D anatomy from 2D standard cardiac views, effectively enabling 3D anatomical reconstruction from limited 2D data. We evaluate our pipeline using synthetically generated data achieving accurate 3D whole-heart reconstructions (peak intersection over union score $> 0.88$) from just two standard anatomical 2D views of the heart. We also show preliminary results using real echo images.

\end{abstract}

\section{Introduction}
\subsection{Motivation and background}
The most common imaging modality for cardiac assessment worldwide is ultrasound (US) \cite{Braga2019} because it is more affordable and safer, and it has higher temporal resolution when compared to other modalities. The reliance on operator skill to acquire high quality standard views is the main limitation of US, particularly for cardiac applications (echocardiography).

Most US systems only have 2D capabilities, and most echo protocols are limited to 2D modes \cite{Robinson2020}. 3D data is desirable both for anatomical and functional understanding of the heart. The heart is a complex 3D structure, and subsequently most features cannot be fully captured within 2D planes. Ventricular and atrial walls are curved surfaces, valve hinges do not sit on a 2D circle but rather a 3D saddle shape, and as a result current protocols require many standard planes to be able to assess most features, while a high quality 3D reconstruction would capture them all. Cardiac function is quantified primarily via tissue motion and blood flow. Blood quantity is volumetric, but usually 2D approximations are used, e.g. 2D left ventricular delineation on a 4 chamber view to compute ejection fraction. This has been commonly used both in classical clinical disease detection and deep learning methods \cite{Upton2022, Ouyang2020}. Tissue motion is analysed using 2D components of the motion e.g. longitudinal, radial, etc., however cardiac motion is complex and combines torsion, vertical motion and displacement, which are fundamentally 3D phenomena. Capturing only 2D views can impede further analysis if the quality of the views is suboptimal from factors such as foreshortening and artifacts. 

3D volumes would allow for more accurate quantification of the cardiac geometries, and in turn allow clinicians to reduce the rate of disease misdiagnosis as a result of the incorrect quantification arising from using 2D data. 3D ultrasound probes are available for clinical use. They are however very prone to artifacting, particularly in moving anatomies such as the heart, and as a result are very rarely used in practice \cite{Nelson2000}.

The ability to quantify cardiac function using standard 2D US systems, without the need for external trackers to define a spatial geometry, can be of great clinical use, but is limited by lack of paired US and native 3D ground truth anatomy.
The aim of this work is to develop a method for reconstructing a full 3D representation of the heart from 2D image views.

\subsection{Related works}
The advent of Deep Learning (DL) has led to development of 3D image reconstruction methods that have been shown to successfully learn shape and structure from partial observations, achieving start-of-the-art (SOTA) performance in natural and medical imaging.

\subsubsection{2D to 3D reconstruction in natural images}
Current SOTA methods using DL allow accurate 2D to 3D reconstruction of objects from one or more RGB images, without the aid of camera positioning calibration information. A popular standardized dataset used to perform comparisons of reconstruction techniques is ShapeNetCore \cite{Chang2015}. ShapeNetCore covers 55 object categories with over 51,300 unique 3D models. In 2019, ShapeNetCore reconstruction SOTA accuracy was achieved by Xie et al. who were able to reach 0.706 Intersection over Union (IoU) using the Pix2Vox++ (PiVox) network \cite{Xie2019}. Two network variations, Pix2Vox++/Fast (PiVox/Fast) and Pix2Vox++/Accurate, or (PiVox/Accurate), were proposed with the PiVox/Fast network being a lighter weight, albeit lower performing, version of the full PiVox/Accurate network. However, PiVox/Fast is still an expensive model to train and perform inference, thus severely limiting the possible reconstruction resolution on currently available single Graphics Processing Units (GPUs).
 
\subsubsection{Ultrasound specific 2D to 3D reconstruction}
A number of US-specific 3D reconstruction algorithms have been developed showing effective performance but have not been applied to cardiac reconstruction. Cerrolaza et al. \cite{Cerrolaza2018} performed fetal skull reconstruction using conditional hierarchical generative network Variational Auto-Encoders (VAEs) and achieved a Dice Coefficient (DC) of 0.91 using three orthogonal US views. In contrast to our goal of performing cardiac reconstructions, their reconstruction target had a generally regular shape which does not deform during imaging.

Prevost et al. \cite{Prevost2018} combine DL with an Inertial Measurement Unit (IMU), using two consecutive frames with optical flow as a channel to the network along with the Euler angles provided by the IMU. On a phantom they achieved a minimum, medium and maximum drift of 1.70, 18.30 and 36.90 mm respectively.

\subsection{Contributions of this study}
No previous work has addressed 3D cardiac shape reconstructions from untracked 2D echo views. As a first step towards this goal, we present an exploratory investigation using synthetic echo data: 1) A pipeline to synthesize 2D echo views with corresponding 3D ground truth; 2) A demonstration that PiVox can produce accurate 2D to 3D reconstructions of realistic synthetic hearts; 3) A simple modification to both PiVox networks which vastly reduces memory and computational expense, while still achieving high reconstruction accuracy. 

\section{Methods}
\begin{figure}[!h]
    \centering
    \makebox[\textwidth]{\includegraphics[width=\linewidth]{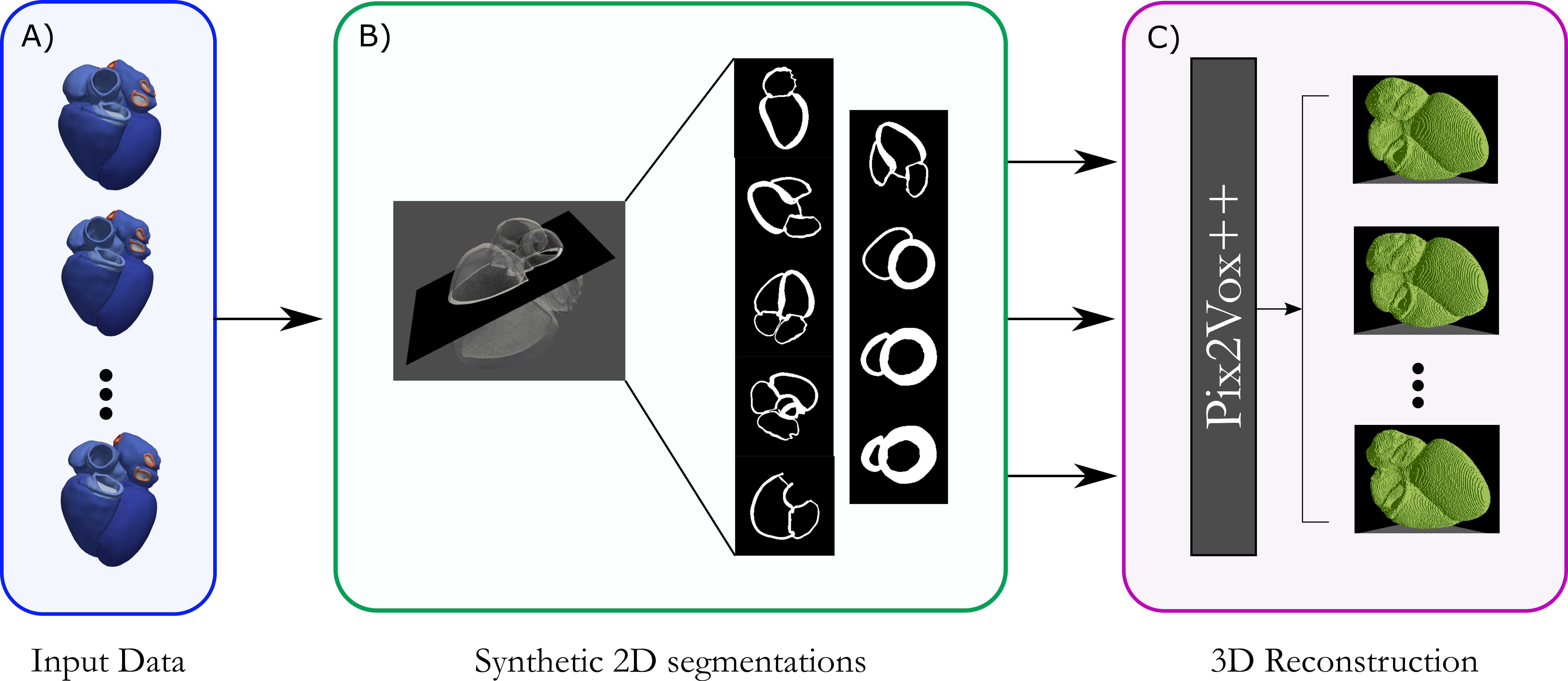}}
    \caption{Proposed pipeline for extracting 2D standard US views from a mesh and performing a reconstruction. A) Initial heart model with tissue labels, B) Mesh slicing with a plane and extracted segmentation-like images and C) Input images being trained/inferenced in the PiVox networks to produce 3D reconstructions.}
    \label{pipleline_fig}
\end{figure}

To explore reconstruction of a 3D heart from 2D data, we used two types of synthetic data: 1) A segmentation dataset, containing binary tissue masks (segmentations) on 2D standard views. Masks are simulated by slicing 3D computational models (which allows using 3D ground truths for evaluation); 2) A synthetic US dataset, containing synthetic 2D standard US views, generated from the tissue masks. We used both for model training and testing. An overview of the proposed pipeline is shown in Figure \ref{pipleline_fig}. 
 
\newpage
\subsection{Synthetic 2D segmentations from 3D heart models}
We obtain synthetic 2D segmentations by slicing 3D mesh models of the heart at standard echocardiographic views. These will be used to asses both how the 2D to 3D reconstruction networks are affected by the large amounts of variation that will be present in real ultrasound images, and also to provide insight into the feasibility of performing training on synthetic data and testing on real ultrasound data. If it can be shown that training using just synthetic data and testing on real data is feasible then it would allow generation of large training datasets and greatly reduce the amount of paired 3D CT/2D ultrasound data required to validate methodologies.

Our slice extraction technique utilizes a combination of the Visualization Toolkit (VTK) \cite{Lowekamp2013} and the PyVista Python packages \cite{daavoo2022}. The mesh data used was a set of 1000 synthetic cardiac meshes created by Rodero et al. \cite{Rodero2021}.

The first step in extracting the standard plane cardiac views requires defining either 3 landmarks for each view to which a plane can be fit, or 2 landmarks and a projection along a vector. We defined these automatically-calculated landmarks based on their relation to various cardiac structures present in the meshes/segmentations (cf. Table \ref{tab:landmarks}). A trained sonographer examined 10 sets of segmentations to confirm the chosen landmarks resulted in suitably realistic extracted slices from our synthetic meshes. 

The left ventricular apex (LVA) was found by performing a ray casting from the centre of mass of the mitral valve to all mesh cell faces in the LV mesh. If there was a pair of intersection points along the ray line (i.e. the ray penetrates both the endo- and epicardial wall) the distance between these intersection points was calculated. The shortest distance was chosen as the LVA, to find the thinnest location of the cardiac wall and to minimize apical foreshortening. 

\begin{table*}[!htb]
\centering
\caption{Ultrasound views and the corresponding landmarks to define these planes.\label{tab:landmarks}}
\begin{adjustbox}{max width=\textwidth}
    \begin{tabular}{|c|c|}
    \hline
   \textbf{Ultrasound View}  & \textbf{Landmarks} \\
   \hline
   Right Ventricular Inflow & Right Ventricle, Right Atrium, Pulmonary Valve \\
   Left Ventricular Parasternal Long-Axis & Left Ventricle Apex, Mitral Valve, Aortic Valve \\
   Parasternal Short-Axis Aortic Valve Level & Left Atrium, Right Atrium, Aortic Valve \\
   Parasternal Short-Axis Mitral Valve Level & Left Ventricle Apex, Whole Heart\\
   Parasternal Short-Axis Papillary Muscle Level & Left Ventricle Apex, Whole Heart \\
   Parasternal Short-Axis Apex Level & Left Ventricle Apex, Whole Heart  \\
   Apical 4 Chamber & Left Atrium, Right Atrium, Left Ventricle Apex \\
   Apical 5 Chamber & Left Atrium, Aortic Valve, Left Ventricle Apex \\
   Apical 2 Chamber & Left Ventricle Apex, Mitral Valve, Right Ventricle\\
   \hline
   \end{tabular}
    \end{adjustbox}
    \label{standard_plane_landmarks}
\end{table*}

\subsection{Synthetic Ultrasound images}
\begin{figure}[!htb]
    \centering
    \makebox[\textwidth]{\includegraphics[width=\linewidth]{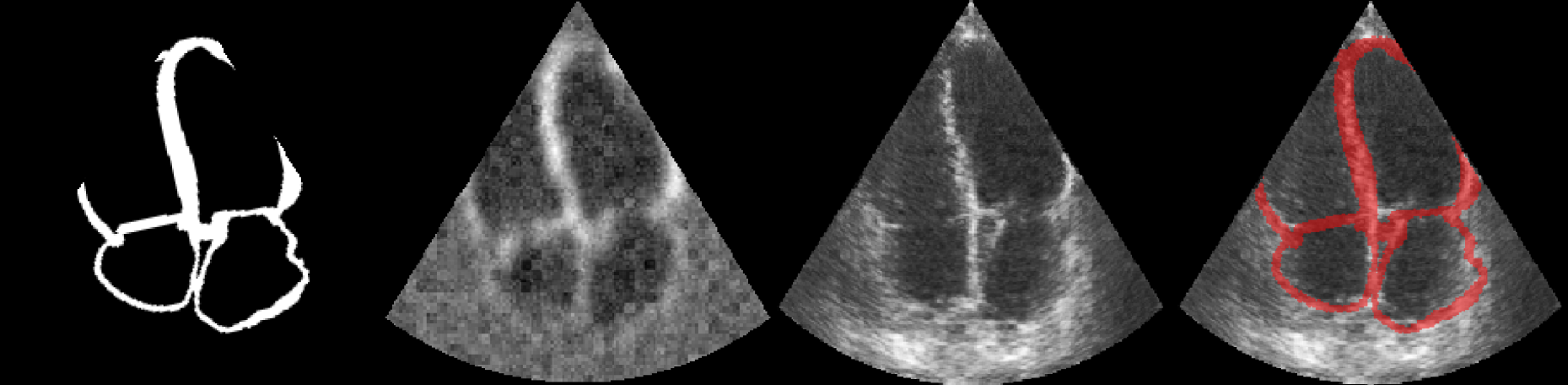}}
    \caption{Example case of data generated at each stage in the process of creating a synthetic ultrasound image from a segmentation mask. From left to right, a) Initial segmentation image b) generated pseudo image c) generated synthetic US d) overlay of input mask on generated image.}
    \label{cyclegan_example}
\end{figure}

Synthetic Ultrasound images (i.e. with real appearance) are generated from the same 3D heart models to study feasibility in realistic echo data while having an exact ground truth. We adapted the technique proposed by Gilbert et al. \cite{Gilbert2021} using the CAMUS dataset \cite{Leclerc2019}. In brief, pseudo-images, i.e. tissue masks corresponding to 2 and 4 chamber images, with noise and Gaussian blurring are used as input to a CycleGAN network \cite{Zhu2017} with unpaired US images to create the final synthetic images, as exemplified in Fig. \ref{cyclegan_example}. 

Producing high quality synthetic US images required small variations in 1) noise parameters 2) sequence of additive noise and blurring, and 3) size of the Gaussian blur kernel. These parameters were changed due to a different resolution input image, and our requirement to not perform any geometric transforms, e.g. changing of input anatomy dimensions.

\subsection{Efficient Pix2Vox++}

The general Pix2Vox++ (PiVox) network architecture is composed of a series of parallel encoder and decoder branches for each input view, before being passed into the fusion and refiner modules. The decoder, in particular, is memory and computationally expensive due to the 3D deconvolutional kernels. This greatly limits the resolution of the 3D volumes that can be used, thus necessitating a more  efficient network with high reconstruction accuracy. It also contains a very large number of learnable parameters, in turn requiring a very large training set. 

Xie et al. proposed a lighter weight, albeit lower performing, variant of the PiVox/Accurate network, referred to as PiVox/Fast, for reducing memory and computational complexity by 1) Using ResNet-18 instead of ResNet-50 2) Decreased de/convolution kernel sizes 3) Removal of the refiner module.

We propose a simple adaptation of the PiVox networks, referred to as E-PiVox, to greatly decrease both memory usage and computational expense, with minimal impact to performance. We achieve this by adding a 3D convolution at the end of the encoder module along the input image dimension to reduce the decoder to a single branch for any number of input images. All information is propagated through the parallel decoder branches in the PiVox networks passes through a compressed latent space, in theory allowing for minimal decreases in performance from having just a single decoder branch rather than multiple branches.

\subsection{2D to 3D reconstruction of synthetic hearts} 
\subsubsection{Implementation details}
The dataset of 1000 cardiac meshes was divided into a 70/15/15$\%$ train/validation/test split and remained unchanged for each of the different training runs. The models were trained with binary segmentation masks, the label maps converted to realistic echo-like images using the CycleGAN network and also on the ShapeNet dataset. All training was performed using Pytorch 1.9.1 \cite{Paszke2019} on an Nvidia RTX 3090 for 200 epochs. The Adam optimizer was used with a $\beta_1$ of 0.9 and a $\beta_2$ of 0.999 \cite{Kingma2015}. The code is available at \href{https://github.com/david-stojanovski/E-Pix2Vox-reconstruction}{https://github.com/david-stojanovski/E-Pix2Vox-reconstruction}

\section{Experiments and Results}
Reconstruction accuracy was assessed using the thresholded Intersection over Union (IoU)~\cite{Xie2019}.

The final reconstruction results of the ShapeNet and cardiac training runs are shown in Tables \ref{shapenet_iou} and \ref{cardiac_iou} respectively. Table \ref{shapenet_iou} shows PiVox/Accurate was consistently the best performing network, however E-PiVox/Fast and E-PiVox/Accurate were consistently within $ 1.2 \%$ and $ 2 \%$ respectively of PiVox/Fast and PiVox/Accurate. This was achieved while being much more memory and computationally efficient. 

\begin{table}[!htb]
\centering
\caption{IoU Comparison of multi-view 2D to 3D object reconstruction on the ShapeNet data set for a $32^3$ resolution.}
\begin{tabular}{|l|c|c|c|c|}
\hline
\textbf{Method}      & \textbf{1 View} & \textbf{2 Views} & \textbf{4 Views} & \textbf{8 Views} \\ \hline
PiVox/Fast                & 0.645           & 0.669            & 0.682            & 0.690            \\ \hline
E-PiVox/Fast              & 0.641           & 0.663            & 0.675            & 0.682            \\ \hline
Percentage diff.(\%) & -0.620          & -0.897           & -1.026           & -1.16            \\ \hline\hline
PiVox/Accurate                & 0.670           & 0.695            & 0.708            & 0.715            \\ \hline
E-PiVox/Accurate              & 0.6687          & 0.6836           & 0.6949           & 0.701            \\ \hline
Percentage diff.(\%) & -0.19           & -1.64            & -1.85            & -1.96            \\ \hline
\end{tabular}
\label{shapenet_iou}
\end{table}

\begin{figure}[!htb]
  \centering
  \includegraphics[scale=0.5]{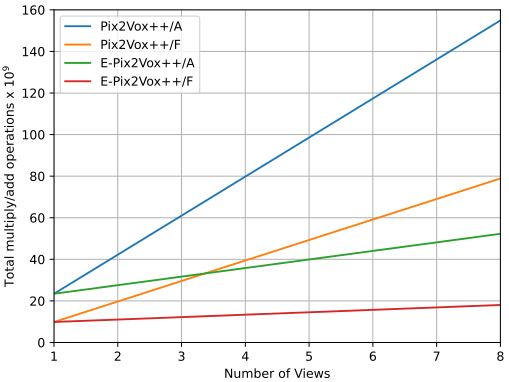}
  \caption{Comparison of the 4 networks and their theoretical total Multiply/Add operations for each number of input views at a $64^3$ resolution.}
  \label{comp_complexity_fig}
\end{figure}

\newpage
It can be seen in Table \ref{shapenet_iou} that our E-PiVox networks was able to come very close to the performance of the much more computationally expensive PiVox networks. The relationship between computational expense and memory usage in regard to number of input views is shown in Figure \ref{comp_complexity_fig} and highlights the importance of efficient network architectures when using a larger number of views, such as in video recordings, or higher resolutions e.g. full resolution CT.

\begin{table}[!t]
\centering
\caption{IoU Comparison of multi-view 2D to 3D object reconstruction on the cardiac data set for a $64^3$ resolution. A2C and A4C refer to atrial 2 and 4 chamber image views respectively, and 9 views denotes all 9 of the previously described standard ultrasound views.}
\begin{tabular}{l|cccc|ccc|}
\cline{2-8}
                                    & \multicolumn{4}{c|}{Binary}                                                                        & \multicolumn{3}{c|}{CycleGAN inferenced}                               \\ \hline
\multicolumn{1}{|c|}{method}        & \multicolumn{1}{c|}{9 views}     & \multicolumn{1}{c|}{A2C}   & \multicolumn{1}{c|}{A4C}   & A2C \& A4C & \multicolumn{1}{c|}{A2C}   & \multicolumn{1}{c|}{A4C}   & A2C  \& A4C \\ \hline
\multicolumn{1}{|l|}{PiVox/Fast}   & \multicolumn{1}{c|}{0.859} & \multicolumn{1}{c|}{0.828} & \multicolumn{1}{c|}{0.841} & 0.818       & \multicolumn{1}{c|}{0.644} & \multicolumn{1}{c|}{0.652} & 0.7     \\ \hline
\multicolumn{1}{|l|}{E-PiVox/Fast} & \multicolumn{1}{c|}{0.86}  & \multicolumn{1}{c|}{0.837} & \multicolumn{1}{c|}{0.85}  & 0.822       & \multicolumn{1}{c|}{0.651} & \multicolumn{1}{c|}{0.66}  & 0.712   \\ \hline
\multicolumn{1}{|l|}{PiVox/Accurate}   & \multicolumn{1}{c|}{0.895} & \multicolumn{1}{c|}{0.85}  & \multicolumn{1}{c|}{0.867} & 0.873       & \multicolumn{1}{c|}{0.673} & \multicolumn{1}{c|}{0.732} & 0.741   \\ \hline
\multicolumn{1}{|l|}{E-PiVox/Accurate} & \multicolumn{1}{c|}{0.903} & \multicolumn{1}{c|}{0.868} & \multicolumn{1}{c|}{0.878} & 0.881       & \multicolumn{1}{c|}{0.674} & \multicolumn{1}{c|}{0.734} & 0.74    \\ \hline
\end{tabular}
\label{cardiac_iou}
\end{table}

Table \ref{cardiac_iou} shows the results when using both binary and realistic ultrasound CycleGAN inferenced images. As shown in Table \ref{cardiac_iou} both E-PiVox networks generally outperformed the PiVox networks in the relative comparisons.The reconstruction accuracy dropped most in areas where the walls of the heart were quite thin, generally in the atrial region. Small errors appearing over the entire structure appear to arise due to errors in the exact location discretization of the voxels.

\begin{figure}[!h]
    \centering
    \makebox[\textwidth]{\includegraphics[width=0.95\linewidth]{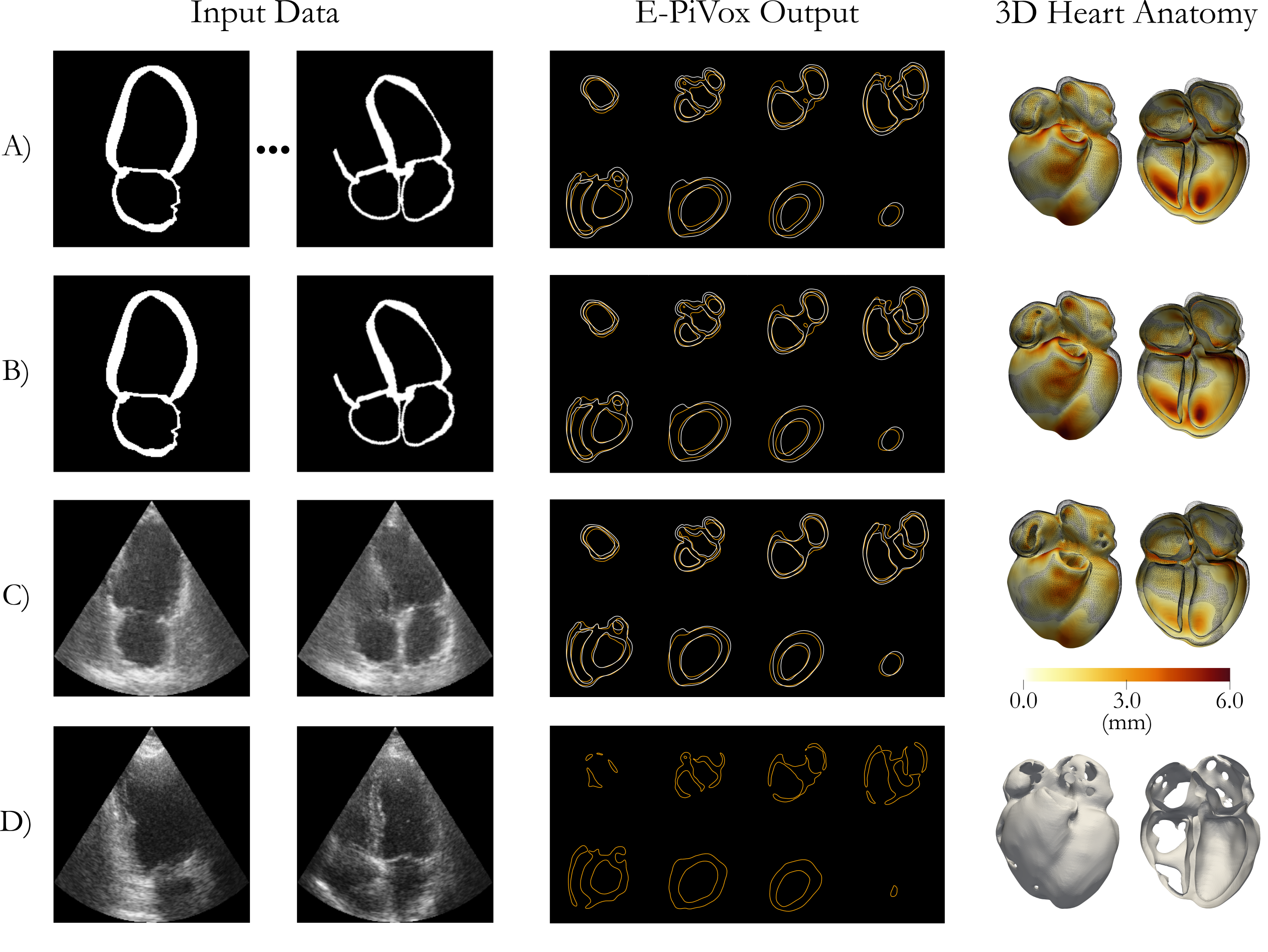}}
    \caption{A) 9 binary views; B) Binary Apical 2 and 4 chamber views; C) CycleGAN inferenced Apical 2 and 4 chamber views; D) Example case from the CAMUS dataset. The same case is shown in A, B and C. The E-PiVox output column shows the ground truth (White) and the network predictions (Orange). The 3D Heart Anatomy column shows the absolute error in relation to the ground truth mesh (Black wireframe).}
    \label{fig:results_3d}
\end{figure}

We present examples of 3D heart reconstructions in Figure \ref{fig:results_3d}, including preliminary results using real 2D echo images from the CAMUS dataset. Synthetically trained reconstructions show close correspondence with the ground truth anatomy (black wireframe). However, reduced number of input views resulted in decreased accuracy and clear non-physiological holes. Such phenomena were exacerbated when using US-like images (see Figure \ref{fig:results_3d}C). Larger differences were observed in the LVA and valve planes for all models.

\newpage
\section{Discussion and Conclusions}
This work provides a proof of concept that a complex geometry like the human heart can be reconstructed with reasonable accuracy from a limited number of standard 2D anatomical views. And that this can be achieved with efficient training and inference. 

 Results in Table \ref{cardiac_iou} show that the PiVox network can successfully reconstruct a full heart with a peak IoU of 0.903 for 9 segmented views, and a minimal decrease in performance to 0.881 for just 2 views. The 0.741 IoU achieved with just 2 synthetic US images represents an acceptable performance given the much more challenging input image data that emulates real-world acquisitions. 
 
 It is important to note that real US images present great variability in contrast and appearance. The preliminary results using real data (panel D in Figure \ref{fig:results_3d}) show reconstructions are possible, albeit needing further improvement. Given a suitably large enough set of real US images, accurate 3D reconstructions could be possible from a small number of standard clinical 2D views. 

The keystone for this contribution is the synthesis of training data (i.e. 2D views) with idealized ground truth 3D reconstructions. This approach addresses the limitations in medical data availability, data privacy, and cost of expert annotation. Preliminary results show promise in the use on real data using models trained with synthetic data, but further research and evidence is needed.

The main limitation of our study is that it is based on the anatomical variability of a synthetic cohort of healthy subjects. As such, the performance could dramatically drop in the presence of disease.

In conclusion, this work demonstrates, in a synthetic workbench, the feasibility of 3D cardiac reconstruction from standard 2D views.

\newpage
\bibliography{my_bib}
\bibliographystyle{splncs04}

\end{document}